\documentclass[aps,prl,twocolumn,floatfix,showpacs]{revtex4-2}

\usepackage[version=3]{mhchem} 
\usepackage{amssymb,amsmath,amsfonts}
\usepackage{graphicx,graphics}
\usepackage{epsfig}
\usepackage{epstopdf}
\usepackage[FIGBOTCAP]{subfigure}
\usepackage{amsfonts}
\usepackage{bm,bbm}
\usepackage{amssymb,amsmath,amsfonts}
\usepackage{mathrsfs}
\usepackage{calc}
\usepackage{epsfig}
\usepackage{float}
\usepackage{color}
\usepackage{epstopdf}
\usepackage{bm,amssymb,amsmath}
\usepackage{graphicx,color}
\usepackage{natbib}

\usepackage{ulem}
\usepackage{parskip}
\newcommand{\Teff}{T_\text{eff}}

\begin{document}
\title{Diffusive contact between randomly driven colloidal suspensions}
\author{Galor Geva$^{(1)}$, Tamir Admon$^{(1)}$, Maayan Levin$^{(1)}$,Yael Roichman$^{(1,2)}$ \footnote{ roichman@tauex.tau.ac.il}}
\affiliation{$^{(1)}$ Raymond \& Beverly Sackler School of Chemistry, Tel Aviv University, Tel Aviv 6997801, Israel \\$^{(2)}$ Raymond \& Beverly Sackler School of Physics and Astronomy, Tel Aviv University, Tel Aviv 6997801, Israel}

\date{\today}
\begin{abstract}
We study the relaxation process of two driven colloidal suspensions in contact, to a joint steady state, similar to the process of thermalization. First, we study a single suspension, subjecting it to random driving forces via holographic optical tweezers, which agitate it to a higher effective temperature. Interestingly, the effective temperature of the suspension, defined by the Einstein relation, exhibits a non-monotonic dependence on the driving frequency. Next, we follow the flux of particles between two such suspensions in diffusive contact, starting from a uniform density and relaxing to a state with zero net particle flux. At high driving frequencies, we show that the density distribution at steady state is determined by equating the ratio of the chemical potential to the effective temperature in both systems, reminiscent of the thermal equilibrium behavior.
\end{abstract}

\maketitle

A dilute suspension of spherical, randomly driven colloidal particles provides a unique platform for investigating non-equilibrium statistical physics. Although clearly in a non-equilibrium state, such systems often exhibit well-defined state functions. 
First, the osmotic pressure, denoted as $\Pi$, coincides with the thermodynamic pressure when defined mechanically, under the condition that interactions between colloidal particles and between them and the boundaries of the system are torque-free \cite{takatoriSwimPressureStress2014, solonPressurePhaseEquilibria2015,solonPressureNotState2015,marini2017pressure,pellicciotta2023colloidal}. 
Second, an effective temperature, denoted as $\Teff$, is well-defined through the generalized fluctuation-dissipation relation (GFDR) when active collisions occur at shorter timescales than the thermal relaxation of the particles to the potential landscape's minima.  
 \cite{Martinez2013,Berut_2014,dieterichSinglemoleculeMeasurementEffective2015,parkRapidprototypingBrownianParticle2020}. 

Notably, these suspensions exhibit an equation of state analogous to that of ideal gases, mirroring the behavior observed in dilute thermal colloidal suspensions: $\Pi=n k_B T_\text{eff}$. Here, $n$ represents the particle number density, $k_B$ is the Boltzmann constant, and $T_\text{eff}$ is an effective temperature that replaces the thermal temperature $T$ in the traditional equation \cite{bonnet-gonnetOsmoticPressureLatex1994,dullensDirectMeasurementFree2006,thorneyworkTwoDimensionalMeltingColloidal2017}. Experiments \cite{ginotNonequilibriumEquationState2015} and simulations \cite{lionOsmosisActiveSolutes2014,takatoriSwimPressureStress2014} have explored deviations from this equation of state as colloidal density increases. Such deviations could arise from the reorganization of the suspension or from memory effects due to complex collision dynamics  \cite{capriniEmergentMemoryTapping2024a}.

Despite the widespread use of effective temperatures  \cite{cugliandoloEffectiveTemperature2011,Puglisi2017temperature}, a significant gap remains in the literature concerning the role of effective temperatures in the relaxation process between two systems in contact as they evolve toward a steady state. For instance, thermal equilibration between systems is independent of their nature: tea cools in air and can also cool by contact with ice. In contrast, far from equilibrium, the effective temperature is often defined in a system-specific manner, rendering the extension of the notion of heat flow between different systems in contact ill-defined. Identifying state functions, such as effective temperature or effective chemical potential, that predict equilibration between driven or active systems in contact is crucial for the generalization of thermodynamics to non-equilibrium conditions.

 	\begin{figure}[t]
	 	\centering
	 	\includegraphics[scale=0.35]{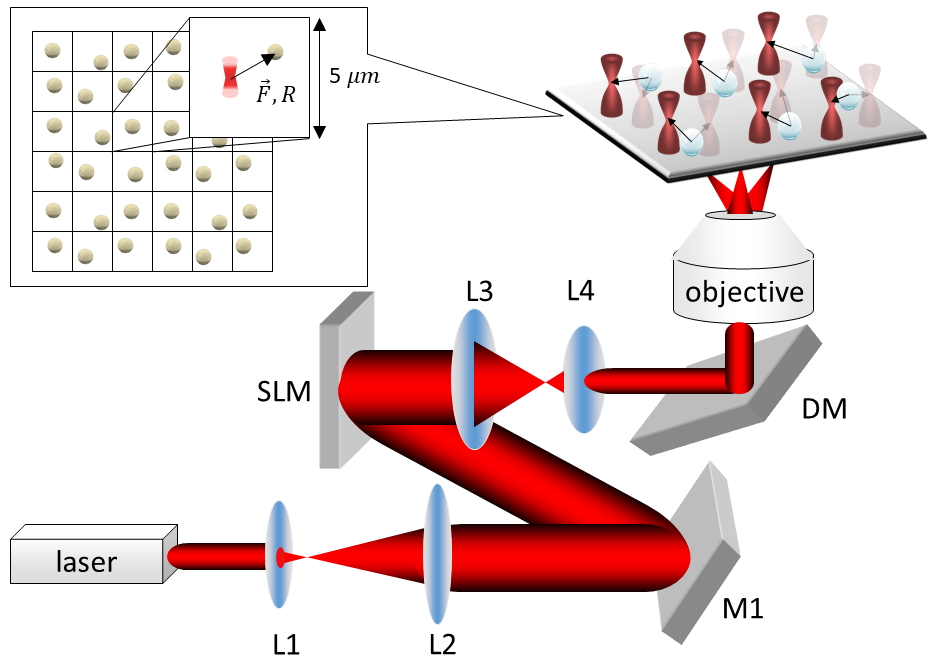}
	 	\caption{Illustration of the HOTs setup and the experiment protocol. Laser light is imprinted by a phase pattern by an SLM. The light is then relayed to the back aperture of an inverted microscope and focused on the sample plane, where it forms a random beam array. The phase mask is changed at a constant switching frequency, thus repeatedly creating new random beam arrays. The transparent beams in the figure represent one array, whereas the bolder beams represent the following beam array. The inset shows a top view of the suspension split into fictitious squares. A uniform beam distribution is achieved by assigning one beam per square.}
	 	\label{RandomTraps}
	 \end{figure}
In this work, we investigate the role of GFDR-based \textit{effective temperature} in the relaxation to a driven steady state of two driven colloidal suspensions in diffusive contact. Our setup allows us to disentangle the influence of driving dynamics from that of the resulting effective temperature on the equilibration process. Specifically, we create a diffusive contact between two driven systems and measure the particle flow between them. We define an effective chemical potential that depends on the effective temperature and demonstrate that no particle flow occurs when the two systems share the same effective temperature and effective chemical potential, even when driven by different driving protocols. Additionally, we show that the ratio of the  generalized effective chemical potential and the effective temperature governs the steady-state particle density in each system when the effective temperature in the two sides is different.

To drive colloidal suspensions in a controlled manner, we apply random optical forces to the colloidal particles and measure the resulting diffusion coefficient $D$ and drag $\gamma$. We calculate the effective temperature using a generalization of the Einstein relation, $T_\text{{eff}} =\gamma D$. We find that the effective temperature is non-monotonic with driving frequency. Specifically, two different driving frequencies can result in the same effective temperature, allowing us to examine the effective temperature's sensitivity to microscopic kinetic details. We find that when well defined, the effective temperature governs the thermodynamics of the driven suspensions independently of the underlying microscopic dynamics. Next, using the ideal-gas-like description of colloidal suspensions, we define an effective chemical potential and show that the equilibration between driven colloidal suspensions in diffusive contact is reached when the ratio of the chemical potential and the effective temperature is equal on both sides, in accord with the entropy maximization rule.
 
We used holographic optical tweezers (HOTs) \cite{dufresne2001computer,Grier2003revolution,Grier2006} to generate multiple randomly placed optical beams (Fig.~\ref{RandomTraps}) and switch their position at constant time intervals $\tau_s=1/f$. Our HOTs setup, previously described in detail \cite{Yevnin_2013}, is based on a Keopsys fiber laser (wavelength, $1083$~nm).  We control the position of the beams by imprinting a phase-only computer-generated hologram on the laser beam using a spatial light modulator (LCOS-SLM, X10468, Hamamatsu) capable of reliably switching its pattern with a frequency $f\leq20$~Hz. We couple the laser pattern on the SLM to the back aperture of a 100x oil immersion objective (NA=1.42) mounted on an Olympus IX71 inverted microscope. The light pattern is then formed in the sample plane. 

Our samples consist of silica particles, $1.50\pm 0.08$~$\mu m$ in diameter (Polysciences, $n_p\approx 1.4$), immersed in a 90$\%$ DMSO  10$\%$ water ($n_m\approx 1.46$) mixture. The higher refractive index of the fluid ensures that the particles are repelled from the laser beams. A quasi-2D suspension of colloidal particles is prepared, approximately 40~$\mu$m high, by placing 8 $\mu$L of the colloidal suspension between a slide and a cover-slip, both passivated with a 10~$\text{w/w}\%$ Bovine serum albumin (Sigma Aldrich) solution.	

We drive the colloidal suspension, mimicking thermal forces in the following manner. An array of 36 randomly positioned beams is projected onto a region of interest (ROI) in the sample plane. To ensure homogeneous driving, we divide the ROI into a fictitious square array with a lattice parameter of $5.0~\pm$~$0.1~\mu m$. Each laser beam is assigned to one square in the lattice, i.e., each beam location is generated randomly within its respective square (Fig.~\ref{RandomTraps}). The random distance between the laser beams and colloidal particles diffusing in the suspension ensures that particles are subjected to forces of varying strengths and directions. The dimensions of the fictitious square array were chosen to obtain a large region of driving without compromising the driving strength (see Fig.~S1, Fig.~S2 for driving force characterization). However, unlike thermal white noise, these tailored fluctuating forces change at constant time intervals. 

The resulting motion of the colloidal particles is a type of Run-and-Tumble motion, where the direction of motion switches completely between runs with no memory or preference for directions. During the runs, the particle is pushed away from the laser beam towards the minimum of the potential landscape by a Gaussian-shaped optical force $\vec{F}_\text{opt}=\frac{A }{\sigma^2}(\Vec{x}-\Vec{y})e^{\frac{-(\Vec{x}-\Vec{y})^2}{2\sigma^2}}$, where $A$ and $\sigma$ are the optical beam's amplitude and width, respectively, $\Vec{x}$ is the position of the particle and $\Vec{y}$ is the position of the beam. The beam's random position $\Vec{y}$ changes randomly at a frequency $f$,  and $\eta$ is a Gaussian white noise of unit variance.

\begin{figure}[t]
	 	\centering
	 	\includegraphics[scale=0.41]{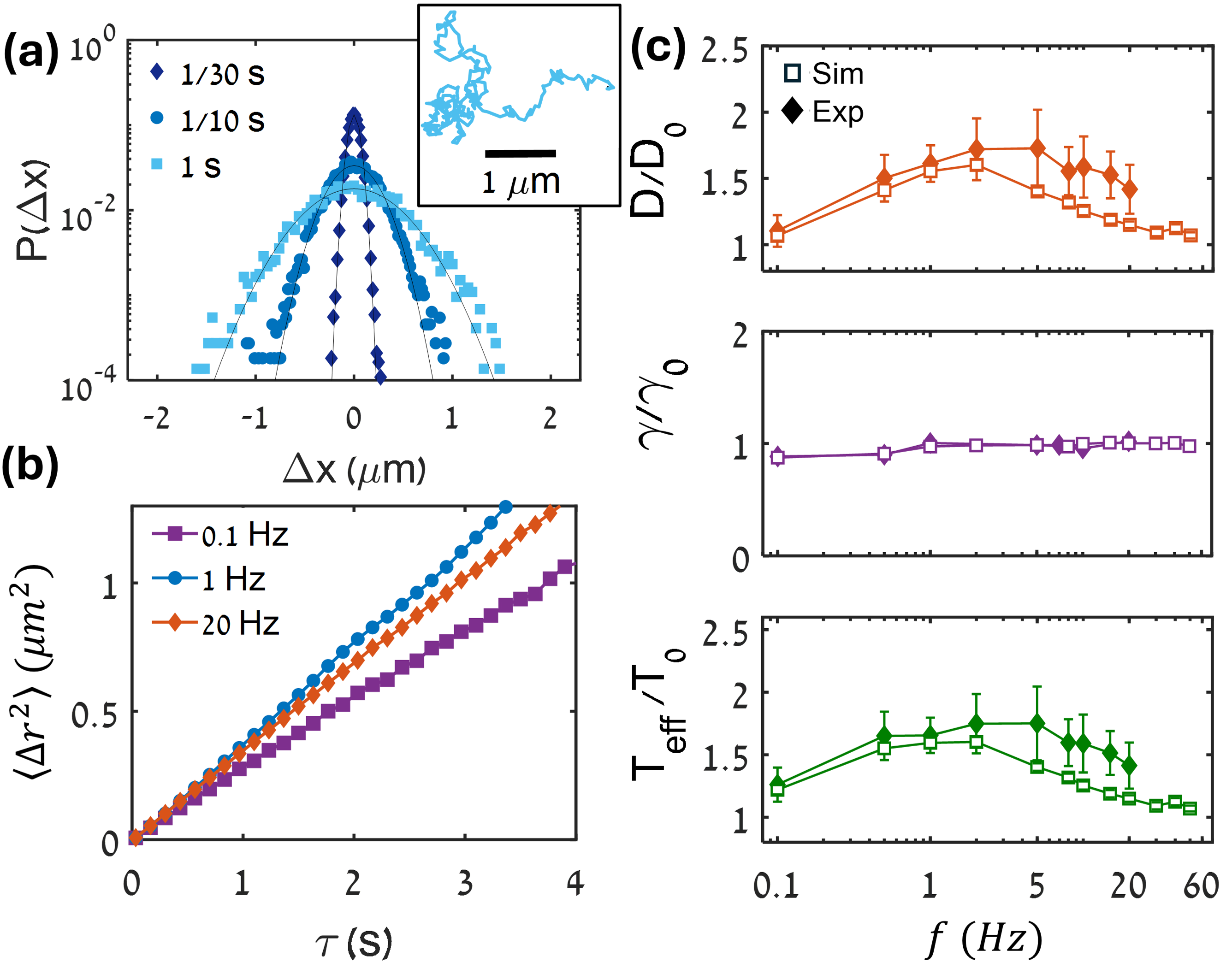}
	 	\caption{Extracting the effective temperature, $T_\text{eff}$, from particle trajectories. a) Probability distribution of particle displacement, at $f=1$~Hz, with lag time $\tau=1/30,1/10,1$~s (symbols) fitted to a Gaussian distribution (solid line). Inset: a typical trajectory of a particle at $f=1$~Hz. b) $\langle\Delta r^2\rangle$ as a function of $\tau$, for experimental measurements of $f=0.1,1,20$~Hz. c) Experimental measurements (filled diamonds) and simulation results (empty squares) of the diffusion coefficient, friction coefficient, and effective temperature normalized by their value in the absence of driving as a function of switching frequency $f$, where $T_0 \approx$ 293 K  is the measured ambient temperature, $D_0 = $ 0.065 $\mu \text{m}^2/\text{s}$  is the diffusion coefficient of a free particle, and $\gamma_0 = 6.18*10^{-8}$ kg/s is the drag coefficient calculated via the Einstein relation.}
	 	\label{fig:TeffD}
	 \end{figure}
  
To characterize the relation between the driving frequency and the resulting effective temperature of the colloidal suspension, we perform a set of experiments in which we change the driving frequency in the range of $0.1~\text{Hz}\leq f \leq 20~\text{Hz}$. We record 10 movies of 5.5~min at 30 fps using a CMOS camera (GS3-U3-2356M, FLIR) for each driving frequency from which we extract particle trajectories using standard particle tracking algorithms \cite{Crocker1996}. A typical trajectory at a driving frequency of $f=1$~Hz is shown in  the inset of Fig.~\ref{fig:TeffD}a 
and  in Fig.~S3. The probability distribution function (PDF) and mean square displacement $\langle\Delta r^2\rangle$ 
of the suspended particles are then calculated (Fig.~\ref{fig:TeffD}a,b).  Since the PDF is Gaussian at $\tau>1$~s and $\langle\Delta r^2\rangle$ increases linearly with lag time, $\tau$, we treat the long-time particle motion as normal diffusion (see Fig.~S4 for more details, with Ref.~\cite{Rahman64}). The diffusion coefficient is then given by $D=\langle \Delta r^2\rangle/4\tau$. 

In Fig.~\ref{fig:TeffD}c (top panel), we plot $D/D_0$ as a function of the driving frequency, where $D_0$ is the diffusion coefficient at room temperature with no driving. Interestingly, $D$ is non-monotonic in $f$, reaching a maximum value at $f\sim2$~Hz. This behavior can be understood by looking at the two limiting cases. At the limit
of very slow driving, the projected pattern barely changes while particles diffuse freely, avoiding the laser
beams (see SM Movie 1). Thus, we expect 
$D/D_0 \simeq 1$. In contrast, at very fast driving, the light pattern changes faster than the relaxation of the particles to the new potential landscape. Therefore, the particles experience a
time-averaged uniform potential in which they diffuse freely, resulting in $D/D_0 \simeq 1$  (see SM Movie 2). The maximum in $D$ emerges at an intermediate frequency where the timescale of pattern switching, $\tau_s$, matches the particle relaxation time, $\tau_r$. When $\tau_s \sim \tau_r$, the system is optimally tuned: particles can fully respond to each new configuration of the optical landscape before it changes again. This condition maximizes the particle response to the time-dependent potential: the particles are displaced significantly by each change in the landscape before it shifts again, while minimizing the time in which the particles are not affected by the optical beams per each cycle. This results in a peak in both diffusivity and effective temperature.

We determine the friction coefficient $\gamma$ of the suspension in the presence of driving by a second set of experiments, in which we translate the suspension slowly through the light pattern. We repeat the experiments for each driving frequency, measuring the average particle velocity (see Fig.~S5 for a detailed description of these experiments). The friction coefficient is given by $\gamma=\gamma_0 v_0/v$, where $\gamma_0$ and $v_0$ represent the friction coefficient and average velocity of the suspended particles without external driving. Figure~\ref{fig:TeffD}c (middle panel) shows the weak dependence of $\gamma$ on the driving frequency.

Following previous studies \cite{palacciSedimentationEffectiveTemperature2010, Martinez2013, Berut_2014, solonPressureNotState2015,bechingerActiveParticlesComplex2016},  we use the Einstein relation to define the effective temperature of the suspension, $k_BT_{\text{eff}} =\gamma D$.
The effective temperature shown in Fig.~\ref{fig:TeffD}c (bottom panel) exhibits a non-monotonic dependence on frequency. This behavior mirrors the trend observed in the diffusion coefficient, as the friction coefficient remains relatively constant across the measured frequency range.

We complement our experiments with Brownian dynamics simulations, numerically integrating the Langevin equation, to extend the range of frequencies and driving strengths examined. In our simulations, particles are subjected to thermal and optical forces. Particles interact via hard core repulsion implemented by the WCA potential. In the first set of simulations, we use $A=35~k_BT$, $T_0=293$~K, $\gamma_0=6.178\cdot10^{-8}$ $\text{kg/s}$, and $\sigma=0.64~\mu$m, similar to the experimental conditions, and collect data from 10 simulations of 10~min duration each to extract $D$ and $\gamma$ as a function of $f$. 

We observe a qualitative agreement between simulations and experiments (Fig.~\ref{fig:TeffD}c). Namely, the diffusion coefficient and the effective temperature peak around $f=2$~Hz, while the friction coefficient barely changes with frequency. A slightly lower effective temperature is observed in the simulations, consistent with an enhancement of driving due to hydrodynamic interactions present in the experiments. This effect is minor since hydrodynamic interactions near the cell floor are suppressed (see Fig.~S6 and discussion, which includes Refs.\cite{sokolovHydrodynamicPairAttractions2011,Nagar_2014,svetlizkySpatialCrossoverFarFromEquilibrium2021}). 

\begin{figure}[t]
	 	\centering
	 	\includegraphics[scale=0.445]{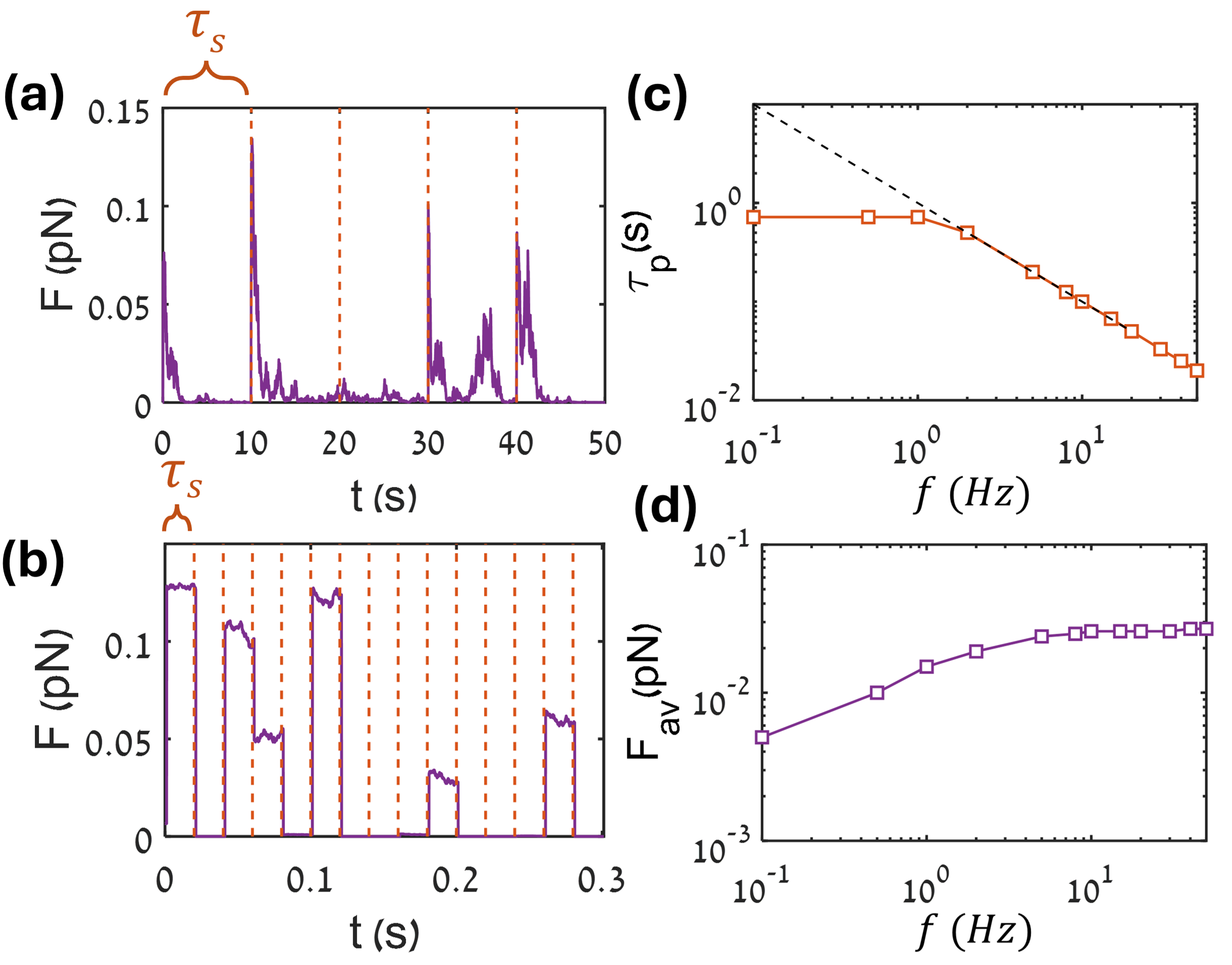}
	 	\caption{Driving characteristics extracted from simulated data with $A= 35~k_BT$. An example of a typical combined force F, (optical and WCA) acting on a single particle as a function of time for a) $f=0.1$~Hz, and b) $f=50$~Hz. c) The persistence particle motion time $\tau_p$ as a function of the switching frequency. d) The time averaged force, F$_\text{av}$,} acting on the particle as a function of the switching frequency.  
	 	\label{fig:TeffF}
	 \end{figure}

We conducted further simulations in which we significantly enhanced the repulsion force of the laser beams (6 fold) and increased the driving frequency beyond experimentally attainable values (up to 60 Hz). This extended the range of examined conditions and allowed us to enhance the effect of optical forces in relation to thermal motion \cite{dieterichSinglemoleculeMeasurementEffective2015, parkRapidprototypingBrownianParticle2020}. 

From the simulated data, we can readily calculate the force acting on each colloidal particle as a function of time, as depicted for $f=0.1$~Hz and $f=50$~Hz in Fig.~\ref{fig:TeffF}a,b, respectively. The initial amplitude of the force is determined by the random distance between the closest beam to each particle. The persistence time of particle motion is given by $\tau_p=\text{min}(\tau_s, \tau_r)$. As the driving frequency increases, $\tau_p$ transitions from $\tau_r$, 0.72~s in our simulations, to $\tau_s$ (Fig.~\ref{fig:TeffF}c).  The average force amplitude acting on the particles increases with frequency and saturates towards a finite value (Fig.~\ref{fig:TeffF}d, Fig.~S7). For Run-and-Tumble particles $k_B(T_{\text{eff}}-T)= v^2\gamma/2\alpha$ where $v$ is the run speed and $\alpha$ is the tumble rate for $t\gg\alpha^{-1}$ \cite{solonPressureNotState2015}. By substituting $v=F_\text{av}\gamma$ and $\alpha^{-1}=\tau_p$ we obtain $k_B(T_{\text{eff}}-T)\sim F_\text{av}^2\tau_p/\gamma$. The coupled effect of the non-linear change in mean force amplitude and persistent time further justifies the non-monotonic change in the effective temperature observed in our experiments and simulations. 

\begin{figure}[t]
	 	\centering
	 	\includegraphics[scale=0.35]{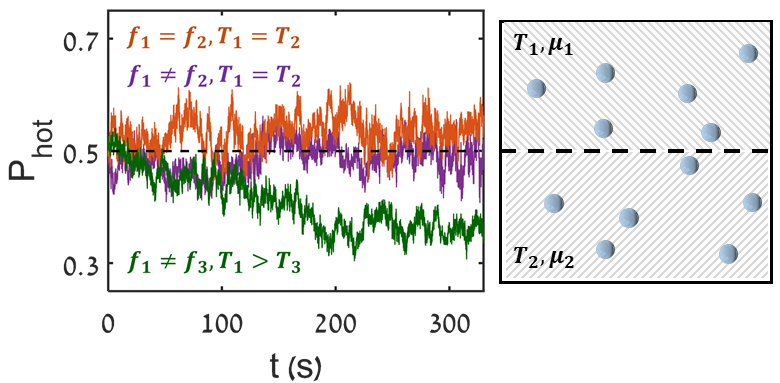}
	 	\caption{Experimental measurement of particle flow between two driven systems in diffusive contact, as illustrated on the right panel. The probability of finding a particle on the region of the higher effective temperature is plotted as a function of time for three typical cases: (1) same driving frequency and same effective temperature (orange), (2) different driving frequency but same effective temperature (purple), and (3) different driving frequency and different effective temperatures (green). Net particle flow is observed only when the effective temperature is different on the two sides. Here, $f_1=1$~Hz, $f_2=2$~Hz, $f_3=0.1$~Hz resulting in $T_1/T_0=T_2/T_0=1.7\pm0.2$ and $T_3/T_0=1.0\pm0.2$. }
	 	\label{fig:Dcontact}
\end{figure}

In equilibrium, the transport of particles between two systems in diffusive contact is determined by the temperature and chemical potential \cite{atkins2023atkins}. The conventional assumption that thermalization is achieved by maximizing entropy with respect to energy and particle flow leads to two requirements, $T_1=T_2$ and $\mu_1/T_1=\mu_2/T_2$, respectively. The chemical potential of dilute colloidal suspensions is given by \cite{1998book}, 

\begin{align}
 \mu=-k_BT\ln{[(\rho_0/\rho)(T/T_0)^{3/2}]},
\end{align}

where $\rho$ is the number density of colloidal particles at temperature $T$, and $\rho_0$ and $T_0$ refer to a reference state. Following the generalization of osmotic pressure in terms of the effective temperature, we generalize the chemical potential of our randomly driven suspensions by substituting $T \to T_{\text{eff}}$ to obtain $\mu_{\text{eff}}\equiv \mu(T_{\text{eff}})$.

With this analogy in mind, our objective is to investigate the particle flow between two regions with different effective temperatures over time. To achieve this, we divide our ROI into two subsystems, each driven independently with a different switching frequency, as depicted in the right panel of Fig.~\ref{fig:Dcontact} and SM movie 3. All experiments start with a uniform distribution of particles. In the diffusive contact experiment, we generated an optical “cage” to
ensure that no particles enter or exit the area of the two subsystems. The cage consists of 56 beams, 1.38 microns apart, forming a square (see SM Movie 4). The addition of optical beams results in weaker driving beams than the previous set of experiments. The effective temperature, which depends linearly on the driving force, is, therefore, smaller in these experiments. We measure the probability of observing a particle in the system at the higher effective temperature region, $P_\text{hot}$, as a function of time. In our control experiment, we drive both sides of the system independently but with the same frequency $f_1=1$~Hz, resulting in zero average current (orange line in Fig.~\ref{fig:Dcontact}). Next, we choose a different frequency $f_2=2$~Hz that drives the suspension to a similar effective temperature $T_{1}/T_0\sim T_{2}/T_0=1.7\pm0.2$. This choice ensures a well-defined effective temperature on both sides \cite{dieterichSinglemoleculeMeasurementEffective2015}. We use $f_1$ as the driving frequency on one side of the system and $f_2$ on the other. While the effective temperature is similar on both sides, the particle dynamics are different. Nonetheless, we observe no net particle flow between the two sides (purple line Fig~\ref{fig:Dcontact}). This result indicates that the effective temperature used here may be a state function of the system and is independent of microscopic dynamics. Finally, we drive the two sides of the system with different frequencies, $f_1=1$~Hz and $f_3=0.1$~Hz, resulting in different effective temperatures, $T_1/T_0=1.7\pm0.2$ and $T_3/T_0=1.0\pm0.2$. Under these conditions particles flow toward the side with a lower effective temperature (green line Fig.~\ref{fig:Dcontact}).
We further demonstrate through simulations that this mechanism is reversible when the conditions in the two contacting baths are switched back and forth (see Fig.~S8).

Since the effective temperature is kept constant by our driving protocol, and particles cannot carry with them effective heat, we expect that the two systems in contact will reach a steady state in which $\mu_{\text{eff}}/k_BT_{\text{eff}}$ is equal on both sides. This condition is held trivially in our first two experiments since the effective temperature $T_\text{eff}$ and particle density $\rho$ are equal from the start. However, for the third experiment, we have $\mu_1/k_BT_3\sim-14.7\pm0.8$ and $\mu_3/k_BT_3\sim-13.3\pm1.3$. These values are equal within experimental measurement, suggesting that in dilute colloidal suspension, the effective temperature defined through the Einstein relation determines the thermodynamic behavior of the suspensions.
\begin{figure}[t]
	 	\centering
	 	\includegraphics[scale=0.34]{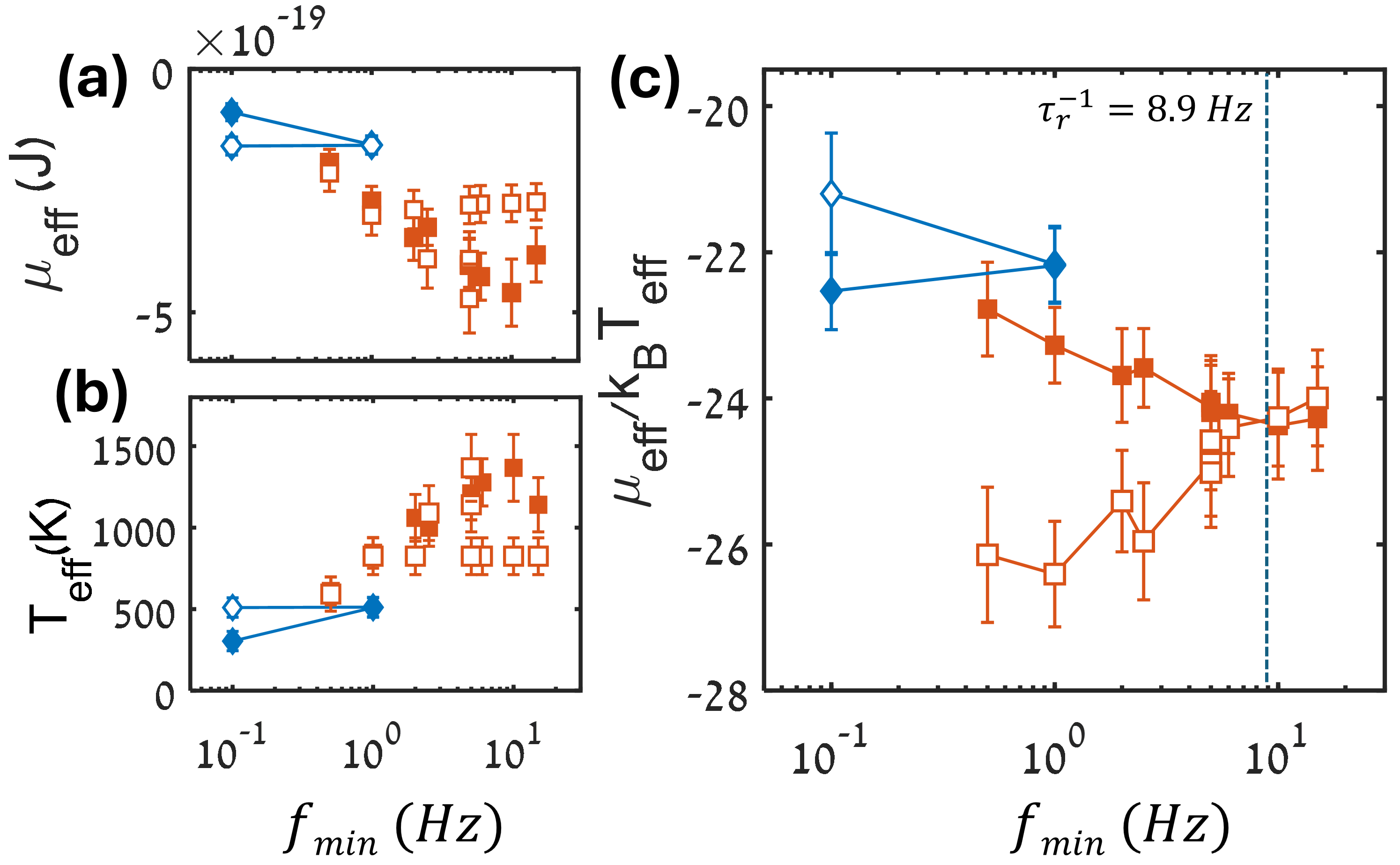}
	 	\caption{Comparison of the thermodynamic variables of two driven suspensions in diffusive contact (empty and filled orange squares for simulations and blue diamonds for experiments). (a) The chemical potential as a function of the minimal switching frequency $f_\text{min}$. (b) The effective temperature as a function of $f_\text{min}$. (c) $\mu_\text{eff}/k_BT_\text{eff}$ as a function of $f_\text{min}$. Simulations were run with $A= 200 k_BT$. Error bars were estimated using the propagated errors of the relevant effective temperatures and steady state densities.}
	 	\label{fig:Thermalization}
   \end{figure}

  Next, we conduct a series of simulations to investigate the spectrum of driving frequencies wherein equilibrium conditions for reaching a steady state apply. Given that the GFDR holds for $\tau_s<\tau_r$, we examine our findings through a plot depicting the lower driving frequency, $f_{\text{min}}$, of the two interacting systems. Specifically, we analyze both systems' effective chemical potential, temperature, and ratio (Fig.~\ref{fig:Thermalization}). We find that as $f_{\text{min}}$ increases, the values of $\mu_\text{eff}$ and $T_{\text{eff}}$ change less with the driving frequency. Moreover, the values of $\mu_\text{eff}/T_{\text{eff}}$ on both sides agree within measurement error at steady-state only when both driving frequencies are sufficiently high. At these frequencies the driven colloidal suspensions behave as equilibrium colloidal suspensions at elevated temperatures.

Our experiment successfully establishes a system for controlling fluctuations in colloidal particles suspended in a fluid. These randomly driven particles exhibit enhanced normal diffusion at longer timescales. Following prior research on driven and active matter, we employed the Einstein relation to define an effective temperature.
Specifically, when the switching time surpasses the average relaxation time of particles within the random driving potential, we ascertain the meaningful thermodynamic implications of this effective temperature. This threshold aligns with the condition previously identified for the temporal fluctuation-dissipation relation's validity in driven and active colloidal suspensions \cite{dieterichSinglemoleculeMeasurementEffective2015, parkRapidprototypingBrownianParticle2020}, as well as in dry driven and active environments \cite{ojhaStatisticalMechanicsGasfluidized2004,boriskovsky2024fluctuationdissipation}.

Our study highlights two key findings in this regime. First, the specific mechanism driving the effective temperature does not influence the relaxation of connected systems toward a steady state. Second, the effective temperature serves a dual purpose: it defines a generalized chemical potential, and together, these two parameters govern the thermalization process in connected systems.
In other words, the steady state of the connected systems is reached at maximum entropy, as in thermal equilibrium. Remarkably, this occurs notwithstanding the continuous investment of work and the concomitant entropy production required to sustain a driven, steady state \cite{krishnamurthy2016micrometre}.

In this study, our focus was on driven dilute colloidal suspensions. We anticipate that our effective chemical potential may prove inadequate in describing the equilibration of systems in diffusive contact in scenarios where the  GFDR is violated. For instance, this could occur in denser driven/active colloidal suspensions \cite{takatoriSwimPressureStress2014,ginotNonequilibriumEquationState2015,capriniEmergentMemoryTapping2024a}, or in systems exhibiting long-wavelength density fluctuations \cite{Shakerpoor2021,gnoliNonequilibriumBrownianMotion2014}. Conversely, a wide array of out-of-equilibrium systems, such as the fluctuations of red blood cells \cite{Ben-Isaac2011}, driven granular gases \cite{cugliandoloEffectiveTemperature2011,ojhaStatisticalMechanicsGasfluidized2004}, collections of bristle robots \cite{boudet2022effective}, and active colloidal crystals \cite{massana2024multiple}, adhere to the  GFDR. This observation suggests that, for these systems as well, the Einstein-based effective temperature may hold thermodynamic significance, for example, in contact where energy and particles are exchanged. 

\begin{acknowledgements}
The authors thank Prof. Nir Gov of the Weizmann Institute of Science, Prof. Haim Diamant of Tel Aviv University, and Dr. Indrani Chakraborty of the BITS-Pilani Goa campus for their insight and fruitful discussions. YR acknowledges support from the Israel Science Foundation (Grant No. 988/17 and 385/21). ML and YR acknowledge support from the European Research Council (ERC) under the European Union’s Horizon 2020 research and innovation program (Grant Agreement No. 101002392).
\end{acknowledgements}


%

\section{Supplemental Information}

\subsection{Calculation of the optical forces}
\begin{figure}[h]
    \centering
    \includegraphics[width=0.5\linewidth]{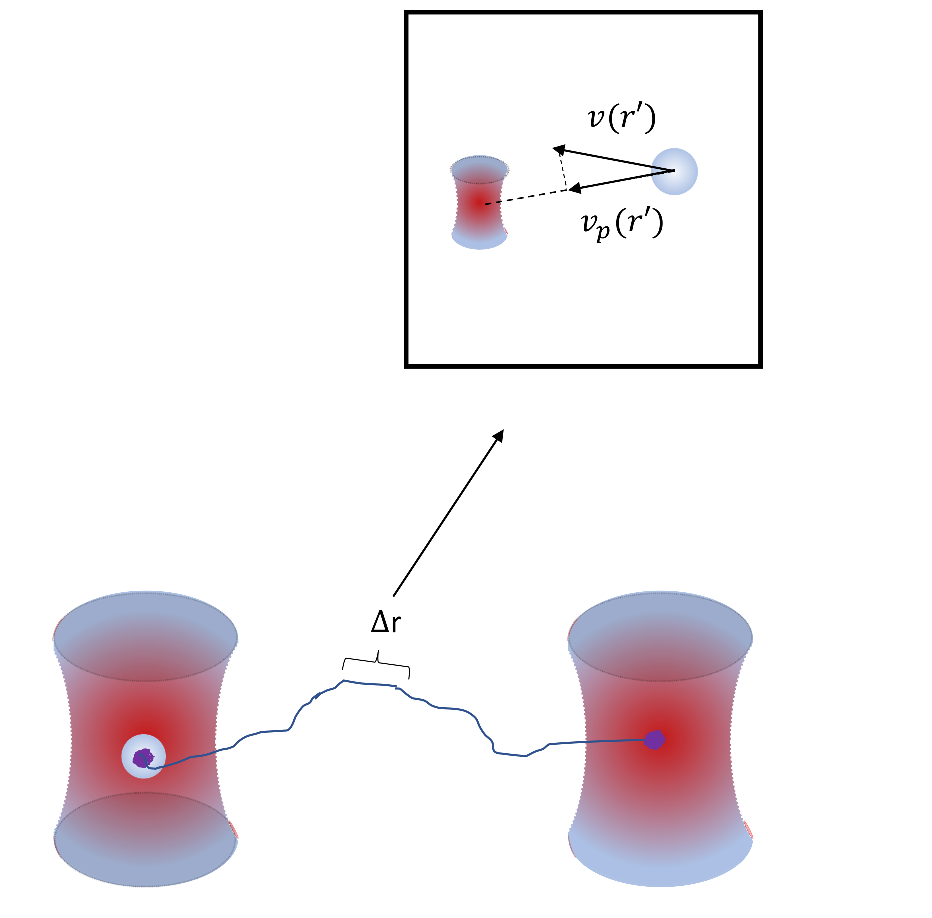}
    \caption{Trajectory of a particle after the trap position is switched. The particle is driven from the former to the current trap position. Inset - the particle's velocity in a specific step and its projection on the distance between the particle and the trap.}
    \label{ping_pong_schematic}
\end{figure}

We calculated the force exerted by an optical beam as a function of distance between the beam and the particle, denoted as $F_\text{opt}(r)$. To this end, we projected a single optical trap in one of two positions, shifted  $1.5 \: \mu m $ apart along the x-axis. We alternated between the two positions at a frequency of $1 \: \text{Hz}$. A single particle is driven back and forth between the two positions of the optical trap (see Fig.~\ref{ping_pong_schematic}). We estimated the velocity of the driven particle as $v(r')=\frac{r(t+\Delta t)-r(t)}{\Delta t}$, where $\Delta t$ is the time between frames ($\Delta t=1/30$ s). The distance between the particle and the trap was defined as
$r'=$$\frac{r(t+\Delta t)+r(t)}{2}$.  We defined the positions of the traps by tracking a single trapped particle and calculating its mean position. The particles were trapped for roughly 5.5 min, i.e., $10^{4}$ frames.

We then calculated the force as $F_\text{opt}(r')=\gamma v_p(r')$, where $v_p$  is the projection of the velocity on the distance between the particle and the trap, and  $\gamma$ is the drag coefficient in the absence of optical forces. 
We set the laser intensity to $10/36=0.28 \: \text{W}$, so that the amplitude of the trap would be similar to the amplitude of the traps in the experiments.

We note that the efficiency of the SLM depends on the phase pattern that is imprinted on it. Thus, we expect the amplitude of a single trap, calculated from fitting $F_\text{opt}(r)$, to be different than the amplitude of a single trap in the random driving experiment. In addition, light intensity losses along the optical pathway are not considered. However, we assume that the width of the Gaussian beam remains unchanged. Therefore, we generate a grid of 36 traps and approximate the amplitude of the traps in the experiment as $ A=k\cdot S^2$, where k is the trap stiffness, calculated from the variance in the position of the trapped particles using the equipartition theorem $\frac{1}{2}k_BT = \frac{1}{2}k\langle\Delta x^2\rangle$, and $S$ is the trap width, obtained by the fitting of $F_\text{opt}(r)$.

The trap stiffness k, calculated using the equipartition theorem, is
$4.5\:\pm\:0.7 \cdot 10^{-7} \: \text{[N/m]}$. The trap width $S$, obtained from fitting $F_\text{opt}(r)$, is $S=0.61 \pm 0.04 \: \mu \text{m}$. We approximate the amplitude of the traps in the experiment as  $A=k\cdot S^2= 41 \pm 8 \: k_BT$.
\begin{figure}[h!]
    \centering
    \includegraphics[width=0.75\linewidth]{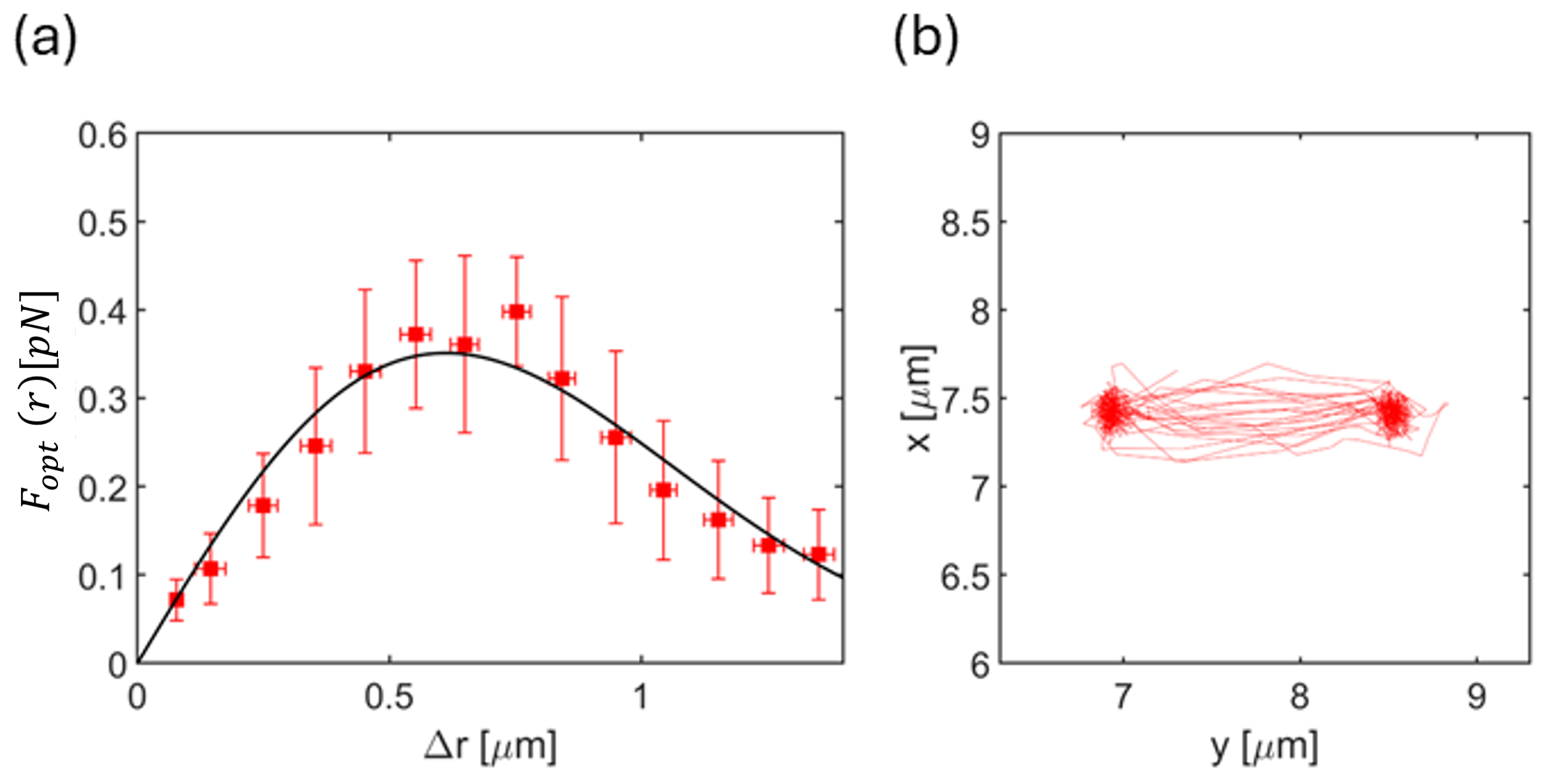}
    \caption{(a) $F_\text{opt}(r)$ in the attractive force setting. The bold black line represents the fitting. The force dependence on the distance is as expected from a Gaussian trap.  The error bars on both the x and y-axis represent the standard deviation of the force and distance, respectively. We find that the force exerted on a particle by an optical trap depends on the distance as $F_\text{opt}(r)= \frac{A}{S^2}r e^{-\frac{r^2}{2S^2}}$, where $S=0.61 \pm 0.04 \: \mu \text{m}$  and $A=87 \pm 7 \: k_BT$. However, we only use the value of $S$ in order to approximate the optical trap in the simulations. (b) Trajectory of a single particle driven back and forth between alternating optical traps.}
    \label{fig:enter-label}
\end{figure}

In the Rayleigh regime, particles are considered to be very small compared to the wavelength of the light. 


The force can be written in terms of the intensity of the light $I$,
\begin{equation}
\vec{F}_{Grad}=\vec{F}_{Grad}=\frac{2\pi n_{m}r^3}{c} \left(\frac{m^2-1}{m^2+2}\right)\nabla I.
\end{equation}
where $m=n_{p}/n_{m}$ is the ratio of refractive indices of the particle ($n_p$) and medium ($n_m$), and $c$ is the speed of light in the medium. By taking into account the refractive indices of water, the colloidal silica particles, and the DMSO+water mixture used in the experiment, and having calculated the trap amplitude in the case of a colloidal particle in an aqueous solution, we 
 insert the respective refractive indices and approximate $\vec{F}_{opt}=35 k_BT.$ 
 
\newpage

\subsection*{Typical trajectory shape}

\begin{figure}[h!]
	 	\centering
	 	\includegraphics[scale=0.3]{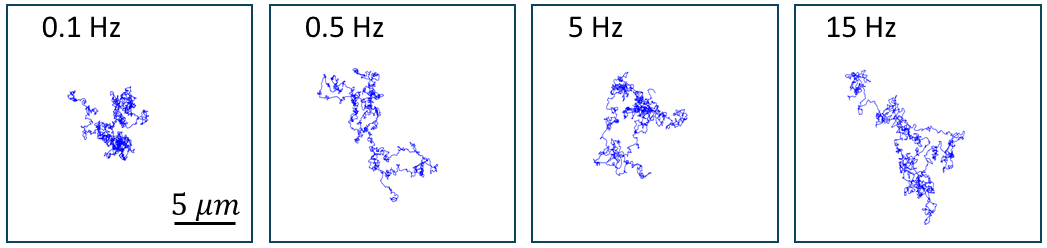}
	 	\caption{Typical trajectories of particles subjected to random repulsive forces switching pattern at a rate of 0.1 Hz, 0.5 Hz, 5 Hz, and 15 Hz. The particles are 1.5 $\mu$m silica particles, suspended in a mixture of 90$\%$ DMSO  10$\%$ water ($n_m\approx 1.46$). A HOTs apparatus creates and switches the laser beam pattern. }
	 	\label{fig:trajectories}
	 \end{figure}

Figure~\ref{fig:trajectories} shows typical trajectories of particles at different driving frequencies. The particles are driven away from the optical beams toward a local minimum of the potential landscape, where they diffuse, unless a new random beam setting is projected driving them toward a new position. The typical trajectories, therefore, show either
directed motion or diffusion. Consequently, at low switching frequencies, particles are confined to more compact areas on average. 

\subsection*{The normal diffusion regime}

 \begin{figure}[h]
    \centering
    \includegraphics[scale=0.25]{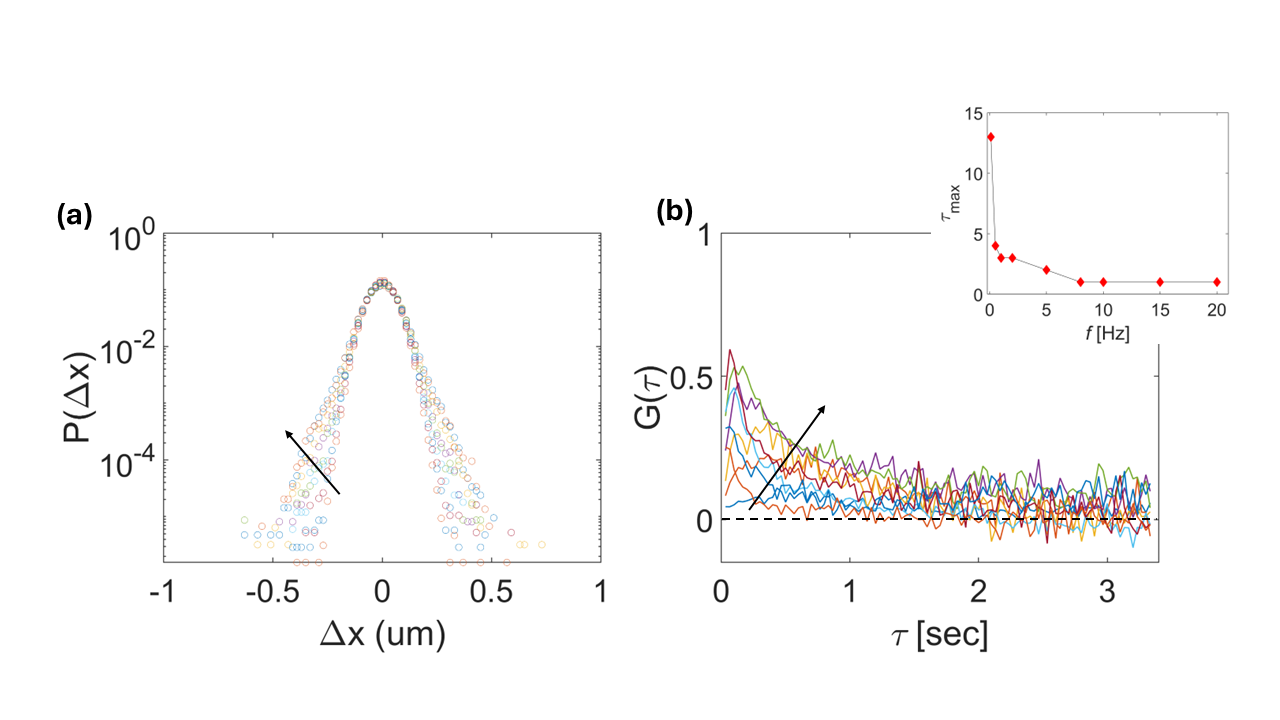}
    \caption{Characterization of the stochastic motion of the driven colloidal particles. a) Probability distribution of the particle’s displacement after the shortest step measured ($\tau = 1/30$~s) for different driving frequencies (various colors). The arrow indicates the trend with increasing driving frequency. b) The non-Gaussian parameter $G(\tau)$ as a function of lag time for different driving frequencies (various colors). The arrow indicates the trend with increasing driving frequency. The inset shows the lag time at which $G(\tau)$ reaches its maximal value, $\tau_\text{max}$, as a function of the switching frequency.}
    \label{fig:NGP}
\end{figure}

The diffusion coefficient of our driven particles is well-defined on timescales much longer than the average duration of the particle's persistent motion. From the particle’s displacement probability distribution at the shortest lag time measured $\tau=1/30$~s, we observe a non-Gaussian propagator, especially at high driving frequencies (Fig.~\ref{fig:NGP}a). However, when plotting the non-Gaussian parameter $G(\tau)=\frac{\overline{\langle \Delta x(\tau)^4\rangle}}{3\overline{\langle \Delta x(\tau)^2\rangle}^2}-1$
\cite{Rahman64} we observe that the particle's motion transitions effectively to normal diffusion for lagtimes larger than $\tau=1$~s (Fig.~\ref{fig:NGP}b). 
  \subsection*{Measuring the friction coefficient from simulation and experiments}
\begin{figure}[h!]
	 	\centering
	 	\includegraphics[scale=0.4]{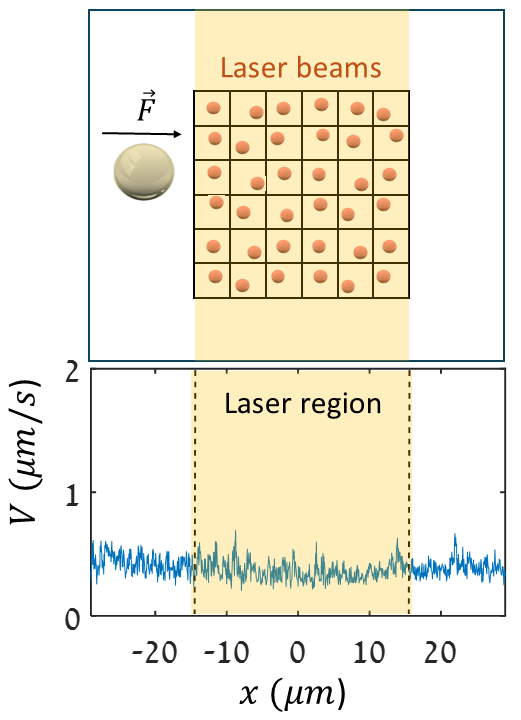}
	 	\caption{Illustration of the friction coefficient measurement scheme (top panel). A typical trace of tracer velocity, at a switching frequency of 0.1 Hz, as it is driven past the fluctuating beam pattern (lower panel).}
	 	\label{fig:friction}
	 \end{figure}

  The measurements of the effective friction coefficient involved translating a colloidal suspension at a constant velocity through a switching optical beam array, at different switching frequencies (top panel of Fig. \ref{fig:friction}). A similar procedure was used in simulations, where a tracer particle experienced a constant drift force in addition to time-varying optical forces. In both cases, the tracer particle's instantaneous velocity was recorded as a function of its position (lower panel of Fig. \ref{fig:friction}). We determined the particle's mobility reduction due to the fluctuating optical potential by comparing its velocity with and without the optical modulation. This reduction signifies an increase in the effective friction, allowing for a quantitative evaluation of how the dynamic optical forces impact the particle's transport.
  
\newpage

  \subsection*{Effect of hydrodynamic interactions on the effective temperature}

\begin{figure}[h]
    \centering
    \includegraphics[scale = 0.55]{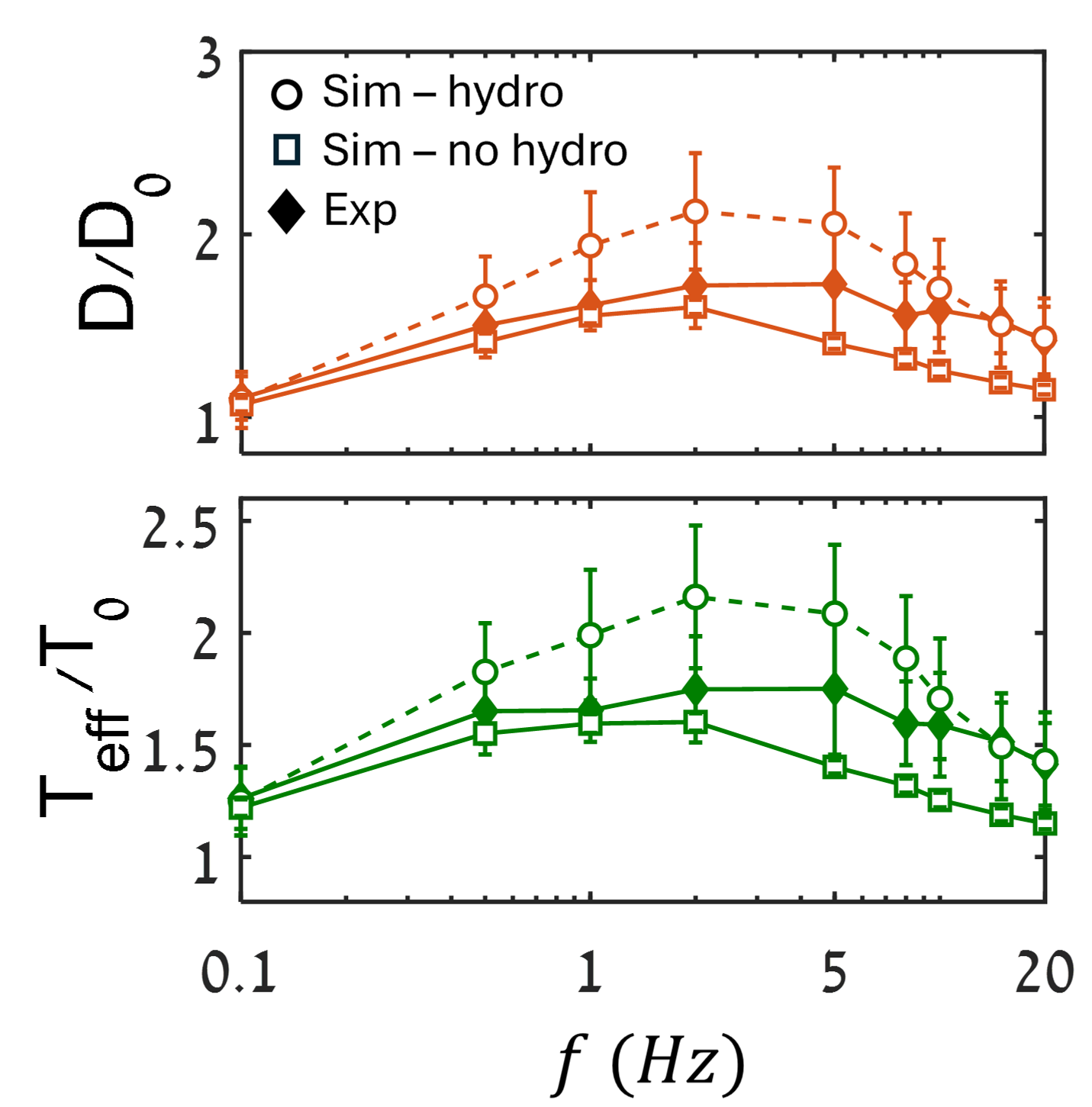}
    \caption{Experimental diffusion coefficients (filled diamonds) are compared with simulation results. The simulations include (empty circles) and exclude (empty squares) hydrodynamic interactions. All data is normalized by the diffusion coefficient and effective temperature without driving and plotted as a function of the switching frequency. }
    \label{fig2_hydro}
\end{figure}

  To examine the effect of hydrodynamic interactions (HI), we performed an additional set of Stokesian dynamics simulations \cite{sokolovHydrodynamicPairAttractions2011,Nagar_2014} in which the Rotne--Prager operator approximates HI and an unbounded 3D fluid suspension is assumed. We expect HI to enhance the agitation of the suspended colloidal particles since they transmit forces moving one particle to other particles. This is observed in Fig.~\ref{fig2_hydro}; the diffusion coefficient of the simulated data with HI is higher than that of simulations without HI. Moreover, the experimental data are in between these limits since the actual HI in the experiment is partially suppressed due to the proximity of the particles to the bottom of the sample \cite{svetlizkySpatialCrossoverFarFromEquilibrium2021}.

\newpage
\subsection*{Relation between potential amplitude and average force}

\begin{figure}[h]
    \centering
    \includegraphics[scale = 0.6]{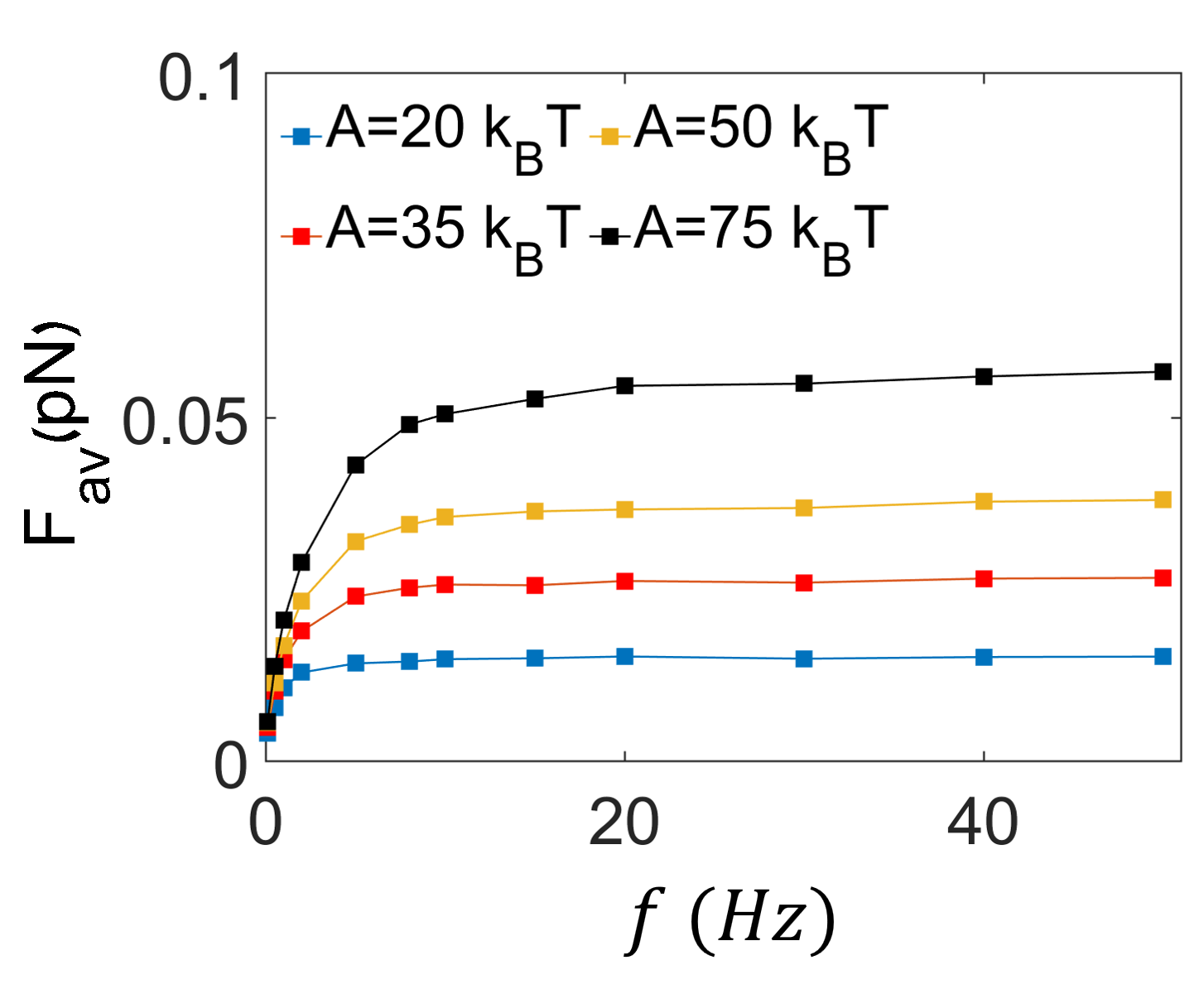}
    \caption{Relation between the optical potential amplitude and the average force experienced by a particle in simulations, $F_{\text{av}}$ as a function of switching frequency for different trap amplitudes. Error bars are omitted since the error,
defined as the standard deviation of $F_{\text{av}}$ between the different particles, is much smaller than the marker
size.}
    \label{F_avg}
\end{figure}

 \begin{figure}[h!]
     \centering
     \includegraphics[width=0.8\linewidth]{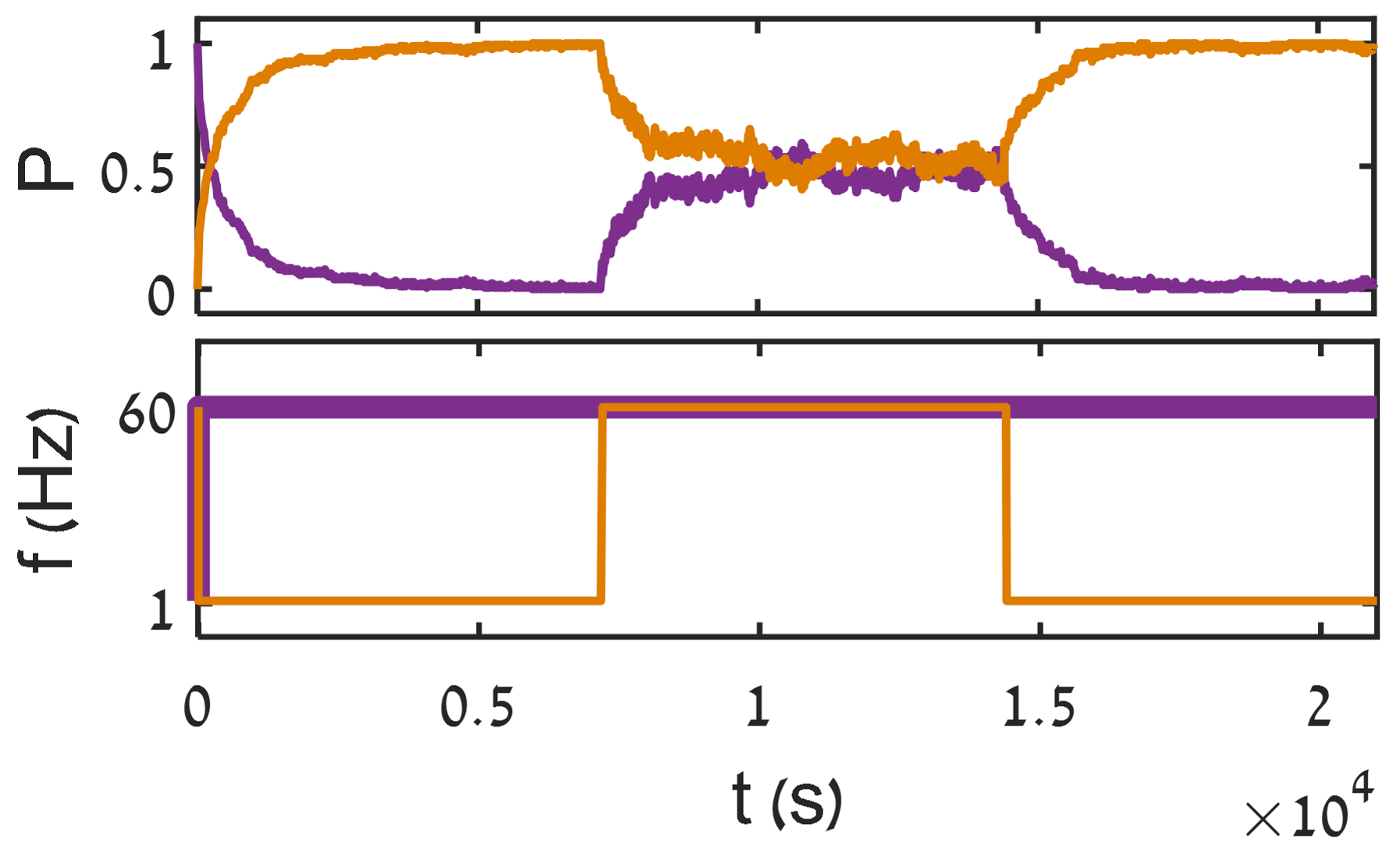}
     \caption{Simulation of reversible process in which particles flow between two driven systems that are in diffusive contact.
     The probability of finding a particle on each side as a function of time is plotted on top. The bottom plot shows the corresponding frequency of each driven system, which being switched from $f=1$~Hz to $f=60$~Hz. Simulations were run with A = $200k_BT$.}
     \label{fig:enter-label}
 \end{figure}

\end{document}